\documentclass[useAMS,usecolumn,usenatbib,usegraphicx]{mn2e}

\title[Ly$\alpha$ haloes at $z=3$]{Diffuse Ly$\alpha$ haloes around Ly$\alpha$ emitters at $z=3$\thanks{Based on data collected at Subaru Telescope, which is operated by the National Astronomical Observatory of Japan.}: Do dark matter distributions determine the Ly$\alpha$ spatial extents?}
\author[Y. Matsuda et al.]{
\parbox[t]{\textwidth}{\vspace{-1cm}
Y. Matsuda,$^{\! 1,2}$\thanks{E-mail: matsuda@astro.caltech.edu} T. Yamada,$^{\! 3}$ T. Hayashino,$^{\! 4}$ R. Yamauchi,$^{\! 4}$ Y. Nakamura,$^{\! 3, 4}$ N. Morimoto,$^{\! 3}$ M. Ouchi,$^{\! 5}$ Y. Ono,$^{\! 5, 6}$ M. Umemura,$^{\! 7}$ and M. Mori\,$^{\! 7}$}\\\\
$^{1}$California Institute of Technology, MS 249-17, Pasadena, CA 91125, USA\\
$^{2}$Department of Physics, Science Site, Durham University, South Road, Durham, DH1 3LE\\
$^{3}$Astronomical Institute, Graduate School of Science, Tohoku University, Aramaki, Aoba-ku, Sendai 980-8578, Japan\\
$^{4}$Research Center for Neutrino Science, Graduate School of Science, Tohoku University, Sendai 980-8578, Japan\\
$^{5}$Institute for Cosmic Ray Research, University of Tokyo, Kashiwa 277-8582, Japan\\
$^{6}$Department of Astronomy, Graduate School of Science, The University of Tokyo, Tokyo 113-0033, Japan\\
$^{7}$Center for Computational Sciences, University of Tsukuba, Tsukuba 305-8577, Japan}

\begin{document}

\date{Accepted 2012 April 20; Received 2012 April 18; in original form 2012 April 4}

\pagerange{\pageref{firstpage}--\pageref{lastpage}} \pubyear{2012}

\maketitle

\label{firstpage}

\begin{abstract}

Using stacks of Ly$\alpha$ images of 2128 Ly$\alpha$ emitters (LAEs) and 24 protocluster UV-selected galaxies (LBGs) at $z=3.1$, we examine the surface brightness profiles of Ly$\alpha$ haloes around high-$z$ galaxies as a function of environment and UV luminosity.  We find that the slopes of the Ly$\alpha$ radial profiles become flatter as the Mpc-scale LAE surface densities increase, but they are almost independent of the central UV luminosities.  The characteristic exponential scale lengths of the Ly$\alpha$ haloes appear to be proportional to the square of the LAE surface densities ($r_{\rm Ly\alpha} \propto \Sigma_{\rm LAE}^2$).  Including the diffuse, extended Ly$\alpha$ haloes, the rest-frame Ly$\alpha$ equivalent width of the LAEs in the densest regions approaches $EW_0\sim 200$\,\AA, the maximum value expected for young ($< 10^7$\,yr) galaxies.  This suggests that Ly$\alpha$ photons formed via shock compression by gas outflows or cooling radiation by gravitational gas inflows may partly contribute to illuminate the Ly$\alpha$ haloes; however, most of their Ly$\alpha$ luminosity can be explained by photo-ionisation by ionising photons or scattering of Ly$\alpha$ photons produced in H\,{\sc ii} regions in and around the central galaxies.  Regardless of the source of Ly$\alpha$ photons, if the Ly$\alpha$ haloes trace the overall gaseous structure following the dark matter distributions, it is not surprising that the Ly$\alpha$ spatial extents depend more strongly on the surrounding Mpc-scale environment than on the activities of the central galaxies.

\end{abstract}


\section{Introduction}

It is believed that gaseous baryons around galaxies (circum-galactic medium or CGM) play an important role in determining the properties of galaxies via gas inflows / outflows across cosmic time \citep[e.g.,][]{2005MNRAS.363....2K, 2006Natur.440..644M, 2009Natur.457..451D, 2010MNRAS.406.2325O, 2011MNRAS.415.2782V, 2011MNRAS.418.1796F}.  Observationally, the CGM have been studied mainly using absorption lines of background QSOs or galaxies \citep[e.g.,][]{2010ApJ...717..289S, 2011ApJ...740...91P, 2012arXiv1202.6055R}.  An alternative window of the CGM is Ly$\alpha$ emitting halo, which should trace the relatively dense gaseous structure around a galaxy \citep[e.g.,][]{2005ApJ...622....7F, 2010MNRAS.407..613G, 2010ApJ...725..633F, 2011ApJ...739...62Z, 2011arXiv1112.4408R, 2012arXiv1203.3803D}.

Diffuse, extended Ly$\alpha$ haloes are likely a generic property of high-$z$ galaxies.  Based on a deep long-slit spectroscopy of a QSO field, \citet{2008ApJ...681..856R} detected diffuse Ly$\alpha$ haloes by averaging the spectra of 27 Ly$\alpha$ emitters (LAEs) at $z=2.6-3.8$, extending to a radius of $r \sim 30$\,kpc above $\sim 10^{-19}$\,ergs\,s$^{-1}$\,cm$^{-2}$\,arcsec$^{-2}$.  \citet{2011ApJ...736..160S} detected more extended ($r \sim 80$\,kpc) Ly$\alpha$ haloes above a similar surface brightness limit by composites of the Ly$\alpha$ images of 92 UV-selected galaxies in three protoclusters at $z$ = 2.3, 2.8 and 3.1.  S11 found that the Ly$\alpha$ haloes can be well fitted by an exponential profile and that the slopes of the Ly$\alpha$ haloes depend weakly on the central Ly$\alpha$ equivalent widths.  

What determines the Ly$\alpha$ halo structure?  The galaxies in S11 are known to reside in galaxy over-dense regions with $\delta_{\rm gal} \sim 4-7$ \citep{2000ApJ...532..170S, 2005ApJ...626...44S, 2011ApJ...740L..31E}, while most of R08's galaxies would reside in blank field.  Moreover, the UV continuum luminosities of the S11's galaxy sample are $\ga 10$ times brighter than those of the R08' s galaxies.  To fill these gaps, we require observations covering larger dynamic ranges of environment and UV luminosity.  Here, we examine the properties of Ly$\alpha$ haloes around LAEs at $z=3.1$ based on our wide-field deep Ly$\alpha$ imaging data.

In this paper, we use AB magnitudes and adopt cosmological parameters, $\Omega_{\rm M} = 0.3$, $\Omega_{\Lambda} = 0.7$ and $h = 0.7$.  In this cosmology, the Universe at $z=3.1$ is 2.0 Gyr old and $1.0\arcsec$ corresponds to a physical length of 7.6 kpc at $z=3.1$.


\begin{table*}
 \centering
 \begin{minipage}{168mm}
  \caption{Properties of composite sub-samples}
  \label{tab1}
  \begin{tabular}{@{}ccccccccccccc@{}}
  \hline
  Sample & N & $\delta_{\rm LAE}^a$ & $NB_c^b$ & $BV^b$ & $NB_c^c$ & $BV^c$ & $EW_0^d$ & $EW_0^e$ & C$_{NB_{c}}^f$ & r$_{NB_{c}}^f$ & C$_{BV}^f$ & r$_{BV}^f$ \\
         &  &  & (mag)  & (mag)  & (mag)  & (mag) & (\AA) & (\AA) & (10$^{-18}$) & (kpc) & (10$^{-18}$) & (kpc) \\
 \hline
 $2.5<\delta_{\rm LAE}^a<5.5$ & 130 & 3.5 & 25.50 & 27.17 & 24.26 & 26.73 & 88 & 183 & 0.7 & 20.4 & 0.3 & 10.3 \\
 $1.5<\delta_{\rm LAE}^a<2.5$ & 273 & 1.9 & 25.36 & 26.94 & 24.49 & 26.47 & 81 & 117 & 1.4 & 13.2 & --- & --- \\
 $0.5<\delta_{\rm LAE}^a<1.5$ & 861 & 0.9 & 25.40 & 26.95 & 24.70 & 26.64 & 79 & 112 & 1.4 & 10.7 & --- & --- \\
  $-1<\delta_{\rm LAE}^a<0.5$ & 864 & 0.1 & 25.35 & 26.97 & 24.77 & 26.68 & 84 & 109 & 1.5 & 9.1 & --- & --- \\
 \hline
 $21<BV^b<25$ & 31 & 0.5 & 23.54 & 24.59 & 22.39 & 23.91 & 50 & 76 & 3.2 & 14.3 & 0.9 & 18.8 \\
 $25<BV^b<26$ & 203 & 0.8 & 24.58 & 25.66 & 23.78 & 25.05 & 51 & 61 & 1.8 & 13.2 & 0.8 & 13.3 \\
 $26<BV^b<27$ & 894 & 0.9 & 25.33 & 26.60 & 24.65 & 26.15 & 61 & 75 & 1.6 & 10.2 & 0.3 & 14.0 \\
 $27<BV^b<29$ & 1000 & 0.9 & 25.53 & 27.59 & 25.00 & 27.39 & 126 & 170 &  0.7 & 11.9 & --- & --- \\
\hline
 Protocluster LBG$^g$ & 24 & 3.7 & 26.19 & 25.31 & 23.05 & 24.18 & 8 & 53 & 3.2 & 18.2 & 5.0 & 7.5 \\
\hline
\end{tabular}
$^a$The LAE (surface) over-densities calculated by smoothing the LAE spatial distributions with a Gaussian kernel of $\sigma$=1.5\arcmin\ (or FWHM$\sim 5$\,$h^{-1}$\,comoving\,Mpc).\\
$^b$The magnitudes (AB) measured with 2\arcsec\ diameter aperture photometry.\\
$^c$The magnitudes (AB) measured with isophotal apertures determined  on the $NB_c$ (Ly$\alpha$) images with a threshold of 31.5\,ABmag\,arcsec$^{-2}$ (or $8 \times 10^{-20}$ erg s$^{-1}$ cm$^{-2}$\,arcsec$^{-2}$).\\
$^d$The rest-frame Ly$\alpha$ equivalent widths from the 2\arcsec\ diameter aperture photometry.\\
$^e$The rest-frame Ly$\alpha$ equivalent widths from the isophotal aperture photometry.\\
$^f$Best-fit parameters assuming a surface brightness profile of $S(r) = C_n exp(-r / r_n)$, where $C_n$ is a constant in units of $10^{-18}$ erg s$^{-1}$ cm$^{-2}$\,arcsec$^{-2}$ and $r_n$ is a scale length.\\
$^g$LBG sample in the SSA22 protocluster from \citet{2003ApJ...592..728S}.
\end{minipage}
\end{table*}

\section[]{Observations and Data Reduction}

We use the same data set as in \citet{2011MNRAS.410L..13M} and \citet{2012AJ....143...79Y}.  Full detail of the observations and data reduction were described in Y12.  Using a narrow-band filter, $NB497$ \citep[CW=4977\,\AA\ and FWHM=77\,\AA ,][]{2004AJ....128.2073H} on Subaru Suprime-Cam \citep{2002PASJ...54..833M}, we observed 12 pointings: Great Observatories Origins Deep Survey-North (GOODS-N), Subaru Deep Field (SDF), three fields in Subaru-XMM Deep Survey (SXDS-C, N, and S) and seven fields around SSA22 (SSA22-Sb1-7).  The SSA22 contains the protocluster at $z=3.09$ \citep{1998ApJ...492..428S, 2000ApJ...532..170S}.  For the SSA22 fields, we obtained broad-band ($B$ and $V$) images in our observing runs.  For GOODS-N, we used archival raw $B$ and $V$-band images \citep{2004AJ....127..180C}.  For the SDF and SXDS fields, we used public, reduced $B$ and $V$-band images \citep{2004PASJ...56.1011K, 2008ApJS..176....1F}.

We reduced the raw data with {\sc sdfred} \citep{2002AJ....123...66Y, 2004ApJ...611..660O} and {\sc iraf}.  For photometric calibration, we used the photometric and spectrophotometric standard stars \citep{1990AJ.....99.1621O, 1992AJ....104..340L}.  We corrected the magnitudes using the Galactic extinction map of \citet{1998ApJ...500..525S}.  We aligned the combined images and smooth with Gaussian kernels to match their seeing to a FWHM of $1.0\arcsec$ or $1.1\arcsec$ depending on the original seeing.  We made $BV$ images [$BV=(2B+V)/3$] for the continuum at the same effective wavelength as $NB497$ and made $NB_{\rm c}$ (continuum subtracted $NB497$) images for emission-line images.  The total survey area after masking low S/N regions and bright stars is $2.4$ square degrees and the survey volume is $1.8 \times 10^6$\,comoving\,Mpc$^3$.  The 1-$\sigma$ surface brightness limits of the $NB_{\rm c}$ images are $0.7-1.2 \times 10^{-18}$\,erg\,s$^{-1}$\,cm$^{-2}$\,arcsec$^{-2}$.


\begin{figure}
\centering
\includegraphics[scale=0.55]{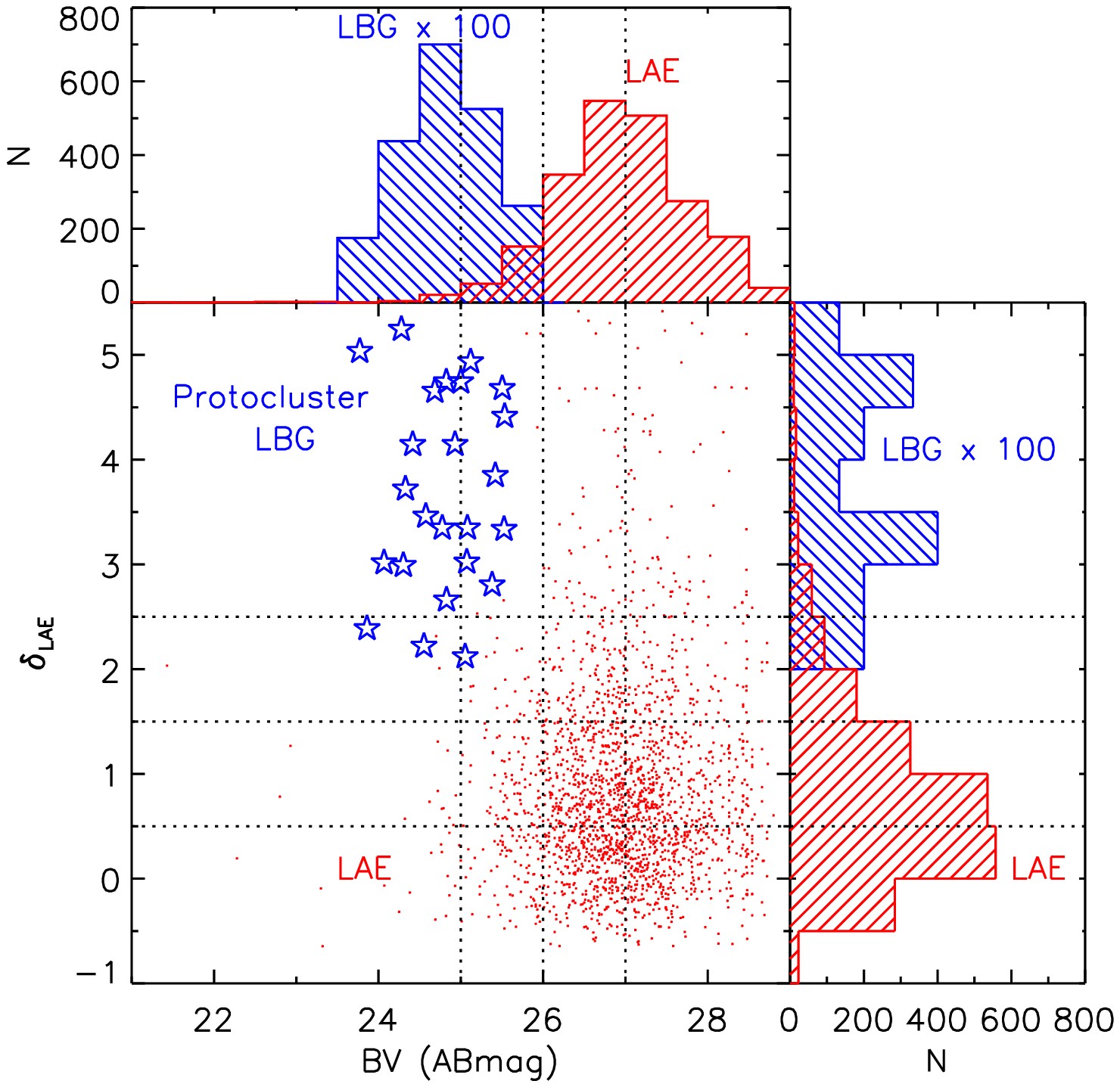}
  \caption{Distribution of $BV$ magnitudes and the surface over-densities of 2128 LAEs at $z=3.1$ (red dots) from \citet{2012AJ....143...79Y} and 24 LBGs at $z=3.1$ in the SSA22 protocluster (blue stars) from \citet{2003ApJ...592..728S}.  The $BV$ is the continuum magnitude at the rest-frame wavelength of $\sim 1220$\AA.  The local over-density of LAEs ($\delta_{\rm LAE}\equiv(\Sigma-\bar{\Sigma})/\bar{\Sigma}$) is calculated by smoothing the LAE sky distributions with a Gaussian kernel of $\sigma$=1.5\arcmin\ (or FWHM$\sim 5$\,$h^{-1}$\,comoving\,Mpc).  The dotted lines represent boundaries between sub-samples for composites.}
\label{fig1}
\end{figure}


\begin{figure*}
\centering
\includegraphics[scale=0.93]{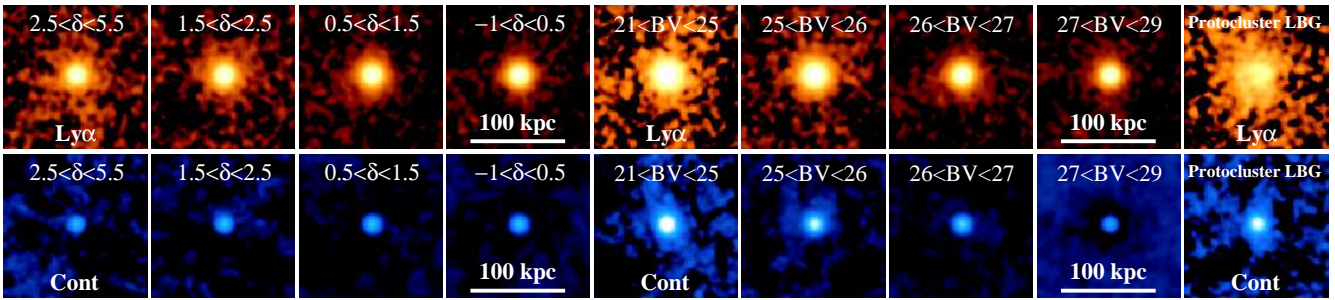}
  \caption{Composite images of $z=3.1$ LAE sub-samples and protocluster LBG sample. The upper panels are Ly$\alpha$ images and the lower panels are $BV$ continuum images.  All the stacked Ly$\alpha$ images show extended Ly$\alpha$ haloes.  The size of the images is $20\arcsec \times 20\arcsec$ ($\sim 150 \times150$\,kpc$^2$).  The white horizontal bars indicate the angular scale of 100\,kpc (physical) at $z=3.1$.  The images are displayed with a logarithmic scale with a range between 26 and 33\,ABmag\,arcsec$^{-2}$.}
\label{fig2}
\end{figure*}

\section[]{Results}

We use a sample of LAEs selected by Y12.  The selection criteria are as follows.

 ${\rm (1)} NB497 < 25.73 ~  {\rm (S/N > 6.6)}$, and, 

 ${\rm (2a)} BV-NB497 > 1.0  ~ \& ~ B-V_c > 0.2$ ($V_c < 26.5$),

 or, 

 ${\rm (2b)} BV-NB497 > 1.3 $ ($V_c > 26.5$), 
 
\noindent where $V_c$ is the line-corrected $V$-band magnitude.  The $B-V_c$ criteria is useful to further prevent the contamination of [O\,{\sc ii}] emitters at $z$ = 0.33 although we cannot apply it to the continuum faint sources, $V_c > 26.5$.  For the robustness of the sample, we also apply an additional constraint that the observed $BV-NB497$ colour must be larger than the 4-$\sigma$ value.  The previous spectroscopic observations show that the contamination of the foreground objects, mostly [O\,{\sc ii}] emitters at $z=0.33$, is at most 1 -- $3\%$, or negligible for the LAE sample \citep{2005ApJ...634L.125M, 2006ApJ...640L.123M, 2012arXiv1203.3633Y}. 

From the total 2161 sample, we exclude 33 LAEs ($1.5\%$ of the whole sample) whose positions are within $20\arcsec$ from the edge of the images.  In Figure~\ref{fig1}, we show $BV$ magnitudes and surface densities of the remaining 2128 LAEs.  The $BV$ magnitude represents the rest-frame UV continuum luminosity near 1220\,\AA.  The surface over-density ($\delta_{\rm LAE}\equiv(\Sigma-\bar{\Sigma})/\bar{\Sigma}$) is derived by smoothing the LAE sky distribution with a Gaussian kernel with a size of $\sigma$ = $1.5\arcmin$ (or FWHM $\sim 5$\,$h^{-1}$\,comoving\,Mpc) as used in Y12, whose scale is close to the median distance between the nearest neighbours of LAEs in the blank fields (SXDS, GOODS-N and SDF).  We confirmed that the result does not change significantly if we use slightly different kernel sizes.  We divide the LAE sample into four sub-samples based on the LAE surface density or the central $BV$ magnitude (see Figure~\ref{fig1} and Table~\ref{tab1}).  We chose the boundaries between the sub-samples as follows.  We first divide the sample into two sub-samples near the peaks of the density or $BV$ magnitude distributions.  Then the dynamic ranges of the high density or $BV$ bright samples are much larger than those of the low density or $BV$ faint samples.  Therefore we further divide the high density or $BV$ bright samples so that sub-samples have basically a dynamic range of $\Delta BV$=1 or $\Delta \delta_{\rm LAE}$=1.  However, as the densest or $BV$ brightest samples have too small numbers of sources, we use larger dynamic ranges ($\Delta BV$=4 or $\Delta \delta_{\rm LAE}$=3) so that they have enough number of sources.  For comparison, we also use a sample of 24 LBGs at $z=3.06-3.13$ (i.e. Ly$\alpha$ redshift range covered by $NB497$) in the SSA22 protocluster from \citet{2003ApJ...592..728S}.


\begin{figure*}
\centering
\includegraphics[scale=0.95]{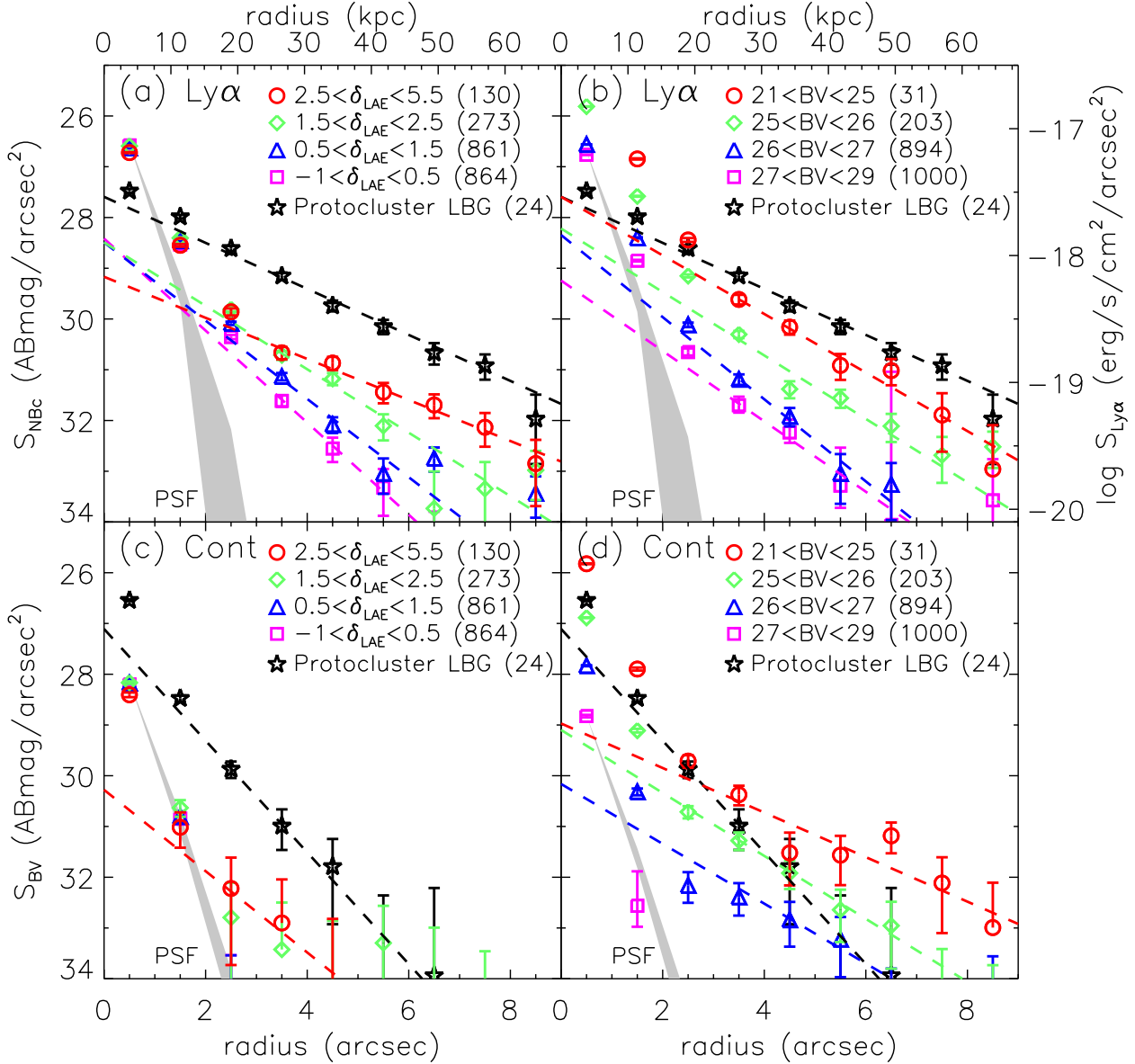}
  \caption{Ly$\alpha$ and continuum surface brightness radial profiles of stacked LAEs and LBGs.  Plots (a) and (b) are Ly$\alpha$ profiles of surface density and $BV$ magnitudes divided sub-samples respectively.  Plots (c) and (d) are UV continuum profiles of surface density and $BV$ magnitudes divided sub-samples respectively.  The grey shades show the range of point spread function (PSF) of the 12 pointings, normalized at the smallest bin of lowest density or faintest UV luminosity sub-samples.  All the Ly$\alpha$ haloes are much larger than the PSF.  For continuum, it is difficult to examine the halo profiles due to their faintness except for the sub-samples with brighter UV continuum.  The dashed lines are fitted exponential profiles, $S(r) = C_n exp(-r / r_n)$, where $C_n$ is a constant and $r_n$ is a scale length.  In order to avoid effect from the PSF and central bright cores, only bins larger than $r=3$\arcsec\ are used for the profile fitting.  The error bar shows a 1-$\sigma$ uncertainty for each bin.}
\label{fig3}
\end{figure*}


\begin{figure*}
\centering
\includegraphics[scale=0.95]{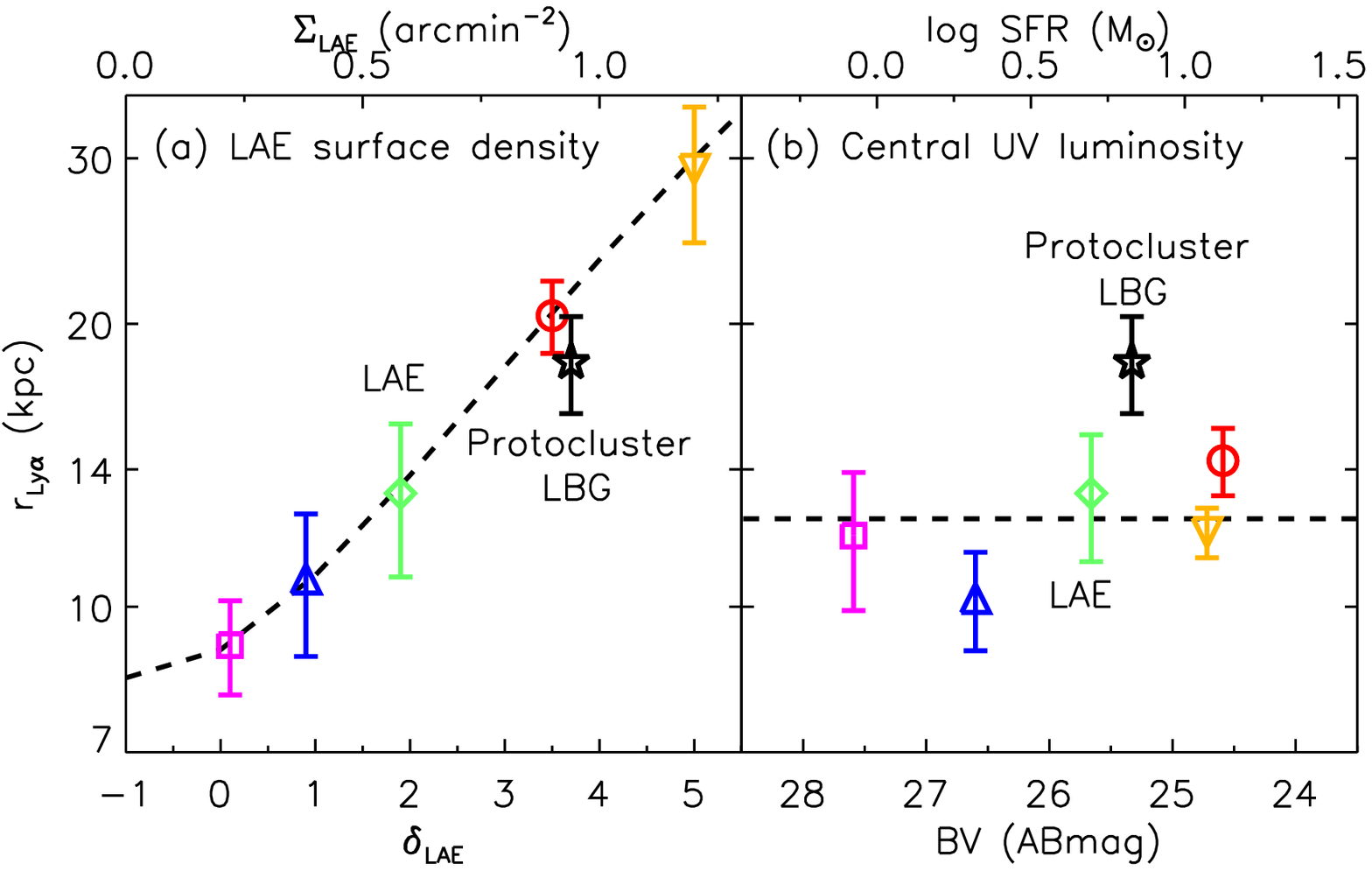}
  \caption{Ly$\alpha$ scale length as a function of (a) LAE surface density and (b) central UV luminosity based on $BV$ magnitude measured with $2\arcsec$ diameter aperture photometry.  The symbols are same as in Figure~\ref{fig3}.  The dashed lines are $r = 0.6 \times (1+\delta_{\rm LAE})^2 + 8.4$\,kpc (left) and $r=12.4$\,kpc (right).  These trends suggest that the spatial extents of the Ly$\alpha$ haloes are determined by the surrounding Mpc-scale environment rather than the central UV luminosities.  The upside down triangles are not main sub-samples shown in Table~\ref{tab1} but parts of sub-samples for checking these trends (see text).  The error bars represent 1-$\sigma$ uncertainties of the exponential profile fitting shown in Figure~\ref{fig3}.  The star-formation rate (SFR) is converted from the $BV$ magnitude without dust attenuation correction \citep{1998ARA&A..36..189K}.}
\label{fig4}
\end{figure*}

Figure~\ref{fig2} shows composite $NB_{\rm c}$ (Ly$\alpha$) and $BV$ (Continuum) images of the LAE sub-samples and the protocluster LBG sample.  We make the composite images as follows.  We cut out $40\arcsec \times 40\arcsec$ (or $300 \times 300$\,kpc at $z=3.1$) images centred at the sources.  We then stack these images with median to erase unrelated foreground or background sources near the sources without masking before stacking.  Median should also be more robust than average if the samples include rare luminous Ly$\alpha$ blobs \citep[LABs, e.g.,][]{2000ApJ...532..170S}.  The 1-$\sigma$ surface brightness limits, estimated from the fluctuations of the outer part of the composite images ($> 10\arcsec$ away from the source), reach $\sim 30.5-32.5$ mag\,arcsec$^{-2}$ or $\sim 0.3-2 \times 10^{-19}$ ergs s$^{-1}$ cm$^{-2}$\,arcsec$^{-2}$, depending on the number of stacked sub-samples.  These depths are equivalent to those achieved with $\sim 30-1000$ nights observation with 8-m class telescopes and is $\ga 10^4$ times fainter than the sky brightness at this wavelength on Mauna kea \citep[e.g.,][]{1987PASP...99..887K}.  All the stacked Ly$\alpha$ images show significant extended Ly$\alpha$ haloes.  The sub-samples with brighter UV continuum show extended UV continuum haloes\footnote{Similar diffuse (and/or clumpy) continuum structure is also seen in giant LABs \citep[e.g.,][]{2007ApJ...667..667M, 2011MNRAS.417.1374P}.}.  For the LAE sub-samples, aperture corrections for $NB_c$ from the $2\arcsec$ diameter aperture to isophotal aperture photometry (above 31.5\,ABmag\,arcsec$^{-2}$ on $NB_c$ images) are $\sim 0.5-1.2$\,mag, while those for $BV$ are $\sim 0.2-0.7$\,mag.  We have confirmed that the results do not change significantly if we use different stacking methods (e.g., 2 or 3-$\sigma$ clipping average).

The composite Ly$\alpha$ and continuum surface brightness profiles are shown in Figure~\ref{fig3}.  We also plot the range of the point spread function (PSF) of the 12 pointings, normalized with the data at the smallest radius for lowest LAE surface density or faintest UV luminosity sub-samples.  The 1-$\sigma$ uncertainty for each bin is estimated from dividing the 1-$\sigma$ surface brightness limit of each composite image with $\sqrt A$, where $A$ is an area of a bin annulus in unit of arcsec$^2$ (i.e., assuming Poisson statistics).  To quantify the shape of the haloes, we fitted them using an exponential profile, $S(r) = C_n exp(-r / r_n)$, where $C_n$ is a constant and $r_n$ is a scale length, by following S11.  In order to avoid contribution from the central galaxies and effect from the PSF, we use only data larger than a radius of 3\arcsec\ for the fitting.  The best-fit parameters are shown in Table~\ref{tab1}.


We check consistency between our work and the S11's work using the protocluster LBG sample.  The Ly$\alpha$ scale length of the protocluster LBG sample ($r=18.2\pm2.2$\,kpc) is similar to that of the S11's full sample ($r=17.5$\,kpc), both of which are derived from median stacking.  In contrast, the UV-continuum scale length of the protocluster LBG sample ($r=7.5 \pm 2.0$\,kpc) seems to be larger than S11's one ($r=3.4$\,kpc).  This difference would come from the different radius ranges used for the profile fittings.  We use data larger than a radius of 3\arcsec\ for the fitting of the haloes.  However, the surface brightness of the UV-continuum profile with $r\ga 3\arcsec$ are fainter than $\sim 31$\,ABmag\,arcsec$^2$ and almost out of the range studied in S11 (as shown in their Figure~7).  If we use data within $r<3\arcsec$ or above $31$\,ABmag\,arcsec$^2$, the UV continuum scale length of the protocluster LBG sample decreases to $r \sim 4$\,kpc and approaches the S11's result.

In Figure~\ref{fig4}, we plot the characteristic exponential scale lengths as a function of the LAE surface over-density and central $BV$ magnitude.  The Ly$\alpha$ scale lengths appear to be roughly proportional to the square of the LAE surface density, $r \propto (1+\delta_{\rm LAE})^2 \propto \Sigma_{\rm LAE}^2$ while there is no clear trend with the central $BV$ magnitude.  We check that the LAE sample in a peak over-density with $\delta_{\rm LAE}=4.5-5.5$ has a Ly$\alpha$ scale length of $r_{\rm Ly\alpha}=29.2\pm4.8$\,kpc and also follows this trend (orange upside down triangle in the left panel).  The Ly$\alpha$ scale length of the $BV$ brightest ($BV=21-25$\,ABmag) sub-sample seems to have slightly higher value than other $BV$ magnitude divided sub-samples.  After excluding 6 sources in $BV=21-24$\,ABmag, which may contain QSOs or AGNs \citep{2002ApJ...576..653S}, the LAE sub-sample with $BV=24-25$\,ABmag has a Ly$\alpha$ scale length of $r_{\rm Ly\alpha}=12.0\pm0.7$\,kpc (orange upside down triangle in the right panel), and is consistent with the average Ly$\alpha$ scale length of the $BV$-magnitude divided sub-samples.  The Ly$\alpha$ scale length of the protocluster LBGs is close to that of the LAE sub-sample with a similar over-density (left panel) but larger than those of the $BV$ magnitude divided sub-samples (right panel).  These results suggest that Ly$\alpha$ scale length is a strong function of the Mpc-scale LAE surface density rather than the central UV luminosity.

\section[]{Summary and Discussion}

Using samples of 2128 LAEs and 24 protocluster LBGs at $z=3.1$, we examine Ly$\alpha$ surface brightness profiles as a function of environment or central UV luminosity.  Our results show that the Ly$\alpha$ scale lengths are roughly proportional to the square of the Mpc-scale LAE surface densities ($r_{\rm Ly\alpha} \propto \Sigma_{\rm LAE}^2$), but are almost independent of the central UV luminosities.  

The apparent profiles of the Ly$\alpha$ haloes should be determined by a combination of gas distribution around galaxies and source(s) of Ly$\alpha$ photons (e.g., R08; S11).  Including the Ly$\alpha$ haloes, the rest-frame Ly$\alpha$ equivalent widths of all the density divided LAE sub-samples are larger than $EW_0\sim 100$\,\AA\ expected for star-forming galaxies with continuous star formation lasting $\ga 10^7$\,yrs (e.g., S11) and that in the over-dense region almost reaches the maximum value of $\sim 200$\,\AA\ for very young ($<10^7$\,yrs) galaxies \citep{1993ApJ...415..580C}.  Therefore, it is possible that Ly$\alpha$ photons formed via shock compression by gas outflows \citep{2006Natur.440..644M, 2012arXiv1203.3633Y} or cooling radiation by gravitational gas inflows \citep{2010MNRAS.407..613G, 2010ApJ...725..633F, 2011arXiv1112.4408R} partly contribute to illuminate the Ly$\alpha$ haloes especially in over-dense environment.  However, more than half of the Ly$\alpha$ luminosity from the haloes can be explained by photo-ionisation by ionising photons or scattering of Ly$\alpha$ photons produced from H\,{\sc ii} regions in and around the central galaxies in any case. 

Although the CGM structure could be modified by gas inflow / outflow processes \citep{2006Natur.440..644M, 2010MNRAS.407..613G}, the overall structure is likely to follow the dark matter distributions \citep[e.g.,][]{2011arXiv1112.1031Y, 2012arXiv1202.6055R}.  From cosmological numerical simulations, \citet{2011ApJ...739...62Z} predicted that Ly$\alpha$ surface brightness profiles are made from a combination of dark matter's one-halo and two-halo contributions with a transition scale of $r\sim 50$\,kpc for a $\sim 10^{11}$\,M$_{\odot}$ halo.  Accounting for this model, our results can be qualitatively interpreted as follows: the observed steeper slope of the Ly$\alpha$ haloes in under-dense regions mainly represent the one-halo term while the flatter slope in over-dense regions is dominated by the two-halo term.  In this case, it is not surprising if the Ly$\alpha$ spatial extents are determined by the surrounding Mpc-scale environment \citep[e.g.,][]{2005MNRAS.364.1327A}.  This interpretation seems to be supported by the known tight links between individually detected LABs and the surrounding large-scale structure: giant LABs have been preferentially discovered in galaxy over-dense regions on Mpc-scales \citep[e.g.][]{2000ApJ...532..170S, 2004AJ....128..569M, 2009MNRAS.400L..66M, 2011MNRAS.410L..13M, 2008ApJ...678L..77P, 2010ApJ...719.1654Y, 2011ApJ...740L..31E}.


This study is the first attempt at investigating Ly$\alpha$ haloes around high-$z$ galaxies in a wide range of environment and central UV luminosity.  However, we lack a sample of UV-bright galaxies with deep Ly$\alpha$ imaging in average to moderate galaxy density environment as shown in Figure~\ref{fig1}.  It would be interesting to test if the Ly$\alpha$ haloes around UV-selected galaxies also show such strong environmental dependence.  In addition to Ly$\alpha$ haloes, our data hint extended continuum haloes at least for UV continuum bright LAE sub-samples.  Although we focus on only Ly$\alpha$ haloes in this paper, further studies of the UV continuum haloes will be important to investigate H\,{\sc ii} regions in the CGM.  Moreover, future $\sim 30$\,deg$^2$ Ly$\alpha$ imaging survey with Subaru / Hyper Suprime-Cam will provide us with $\sim 10$\,k LAE sample (per redshift slice) at $z=2-7$ and enable us to study the redshift evolution of the CGM structure via Ly$\alpha$ haloes. 

\section*{Acknowledgments}

We thank the anonymous referee for the useful comments which significantly improved the clarity of the paper.  We also thank Chuck Steidel and Ian Smail for support and encouragement.  We also thank Tom Theuns, John Stott, Joop Schaye, Zheng Zheng, Michael Shull, Brian Siana, Naveen Reddy, Gwen Rudie, Ryan Trainor and Jean-Rene Gauthier for discussions and comments.  YM acknowledges support from JSPS Postdoctoral Fellowship for research abroad and STFC.

\label{lastpage}

\end{document}